\documentclass[aps,prl,twocolumn,showpacs,superscriptaddress,groupedaddress]{revtex4-1}   	

\usepackage{graphicx}				
\usepackage{amsmath}
\usepackage{amssymb}
\usepackage{dsfont}
\usepackage{dcolumn}	
\usepackage{color}							
\usepackage{bm}

\DeclareMathAlphabet{\mathpzc}{OT1}{pzc}{m}{it}

\newcommand{\ud}{\mathrm{d}}
\newcommand{\Z}{\mathds{Z}}

\begin{document}

\title{Defect in the Joint Spectrum of Hydrogen due to Monodromy}
\author{Holger R. Dullin} \affiliation{ School of Mathematics and Statistics, University of Sydney}
\email{holger.dullin@sydney.edu.au}

\author{Holger Waalkens} \affiliation{Johann Bernoulli Institute for Mathematics and Computer Science, University of Groningen}
\email{h.waalkens@rug.nl}
\date{\today}

\begin{abstract}
In addition to the well known case of spherical coordinates the hydrogen atom  separates in three further coordinate systems.
Separating in a particular coordinate system defines a system of three commuting operators.
We show that the joint spectrum of the Hamilton operator, 
and the $z$-components of the angular momentum and quantum Laplace-Runge-Lenz vectors  
obtained from separation in prolate spheroidal coordinates 
has quantum monodromy for energies sufficiently close to the ionization threshold. This means that one cannot  globally assign 
quantum numbers to the joint spectrum.
Whereas the principal quantum number $n$ and the magnetic quantum number $m$ 
correspond to the Bohr-Sommerfeld quantization of 
globally defined classical actions a third quantum number cannot be globally defined because the third action is globally multi valued.

\end{abstract}

\maketitle

What  could possibly be said about the Hydrogen atom that is new?
The hydrogen atom is conceivably the best studied system 
in quantum mechanics, alongside its classical counterpart  the Kepler problem  in classical mechanics.
These systems are of
paramount importance for our fundamental understanding of atomic physics and astronomy, respectively. 
Using their separability in spherical coordinates these systems are solved in any introductory physics course. 
Separable systems are special examples of integrable systems. 
A  \emph{quantum integrable system} (QIS) is a collection of $f$ independent commuting operators
$\mathcal{H} = (\hat H_1, \dots, \hat H_f)$ where say $\hat H_1$ is the Hamilton operator. 
In addition we require that the classical limits  $ (H_1, \dots, H_f)$ of these operators 
have pairwise vanishing Poisson bracket, and thus constitute a Liouville integrable system
with each $H_i$ a constant of motion (or  integral for short).
Geometrically this means that $(H_1, \dots, H_f)$ defines the \emph{energy momentum map},  from the $2f$-dimensional classical phase space to $ \mathbb{R}^f$ and  that the pre-image of a regular value of this map is a union of $f$-dimensional tori (if compact) in the neighbourhood of which one can construct action angle variables following the Liouville-Arnold theorem \cite{Arnold78}.  
The semiclassical Bohr-Sommerfeld quantization of actions shows that the joint spectrum $ ( \lambda_1, \dots, \lambda_f) \in \mathbb{R}^f$ 
where $\hat H_i \psi = \lambda_i \psi$ for $i = 1, \dots, f$ with joint eigenstate $\psi$, locally has the structure of a lattice $\mathbb{Z}^f$ and can hence be locally labelled by quantum numbers. 

Due to defects this local lattice may not be extendable to a global lattice and hence a global assignment of quantum numbers  to quantum states may be impossible \cite{SadovskiiZhilinskii99,VuNgoc99,Zhilinskii2006}. 
This is the quantum mechanical manifestation of an obstruction to the global construction of action angle variables referred to as Hamiltonian monodromy and first introduced by Duistermaat \cite{Duistermaat80} and then studied quantum mechanically jointly with Cushman \cite{CushDuist88}. 
Now many examples of  (quantum) monodromy are known, see, e.g., \cite{Zhilinskii2011} and the references therein, 
and generalisations have been discovered \cite{Nekhoroshev2006,Sadovskii2007164}. Quantum monodromy explains, e.g., problems in assigning rovibrational spectra of molecules \cite{CDGHJLSZ04,Child2008,Assematetal2010} 
or electronic spectra of atoms in external fields \cite{CushmanSadovskii99,EfstathiouSadovskiiZhilinskii2007}
and it provides a mechanism for excited-state quantum phase transitions \cite{Cejnaretal2006,Caprio20081106}.
The generalization of monodromy to scattering systems leads to similar defects in the lattice of transparent states
in planar central scattering \cite{DullinWaalkens08}. 
It has been shown that monodromy can also play a role in spatiotemporal nonlinear wave systems \cite{Sugnyetal2009}.
Dynamical manifestations of monodromy have recently been studied in \cite{Delos14}.

Both the Hydrogen atom and the Kepler problem have the property that they can be separated in four coordinate systems:
spherical, parabolic, prolate spheroidal (which contains the first two as singular limiting cases) 
and sphero-conical coordinates, see, e.g., \cite{Cordani2003}. 
Each separating coordinate system gives a set of different separation constants which in turn define different sets of three commuting operators.  
In the case of spherical coordinates this gives the very well known QIS $\mathcal{H} = (\hat{H}, \hat{\mathbf{L}}^2, \hat{L}_z )$ which has the
joint spectrum $(-1/(2n^2),\ell(\ell+1),m)$ with quantum numbers $n=1,2,3,\ldots$, $l=0,1,2,\ldots,n-1$ and  $m=-l,-l+1,\ldots,l$.
Here $\hat{L}_z$ is the  $z$-component of the angular momentum operator $\hat{\mathbf{L}}$ and,
as in the rest of this paper, we use atomic units.
In the case of separation in prolate spheroidal coordinates a different QIS is obtained as
$\mathcal{G} = (\hat{H}, \hat{G}, \hat{L}_z)$ with  
$ \hat{G} = \hat{\mathbf{L}}^2 + 2 a\, \hat{e}_z$ 
 where $\hat{e}_z$ is the $z$-component of the quantum Laplace-Runge-Lenz or eccentricity vector 
and the positive parameter $a$ is half the distance between the focus points of the prolate spheroidal coordinates, 
see, e.g.,  \cite{Morse53}.

\begin{figure}
\includegraphics[width=\columnwidth]{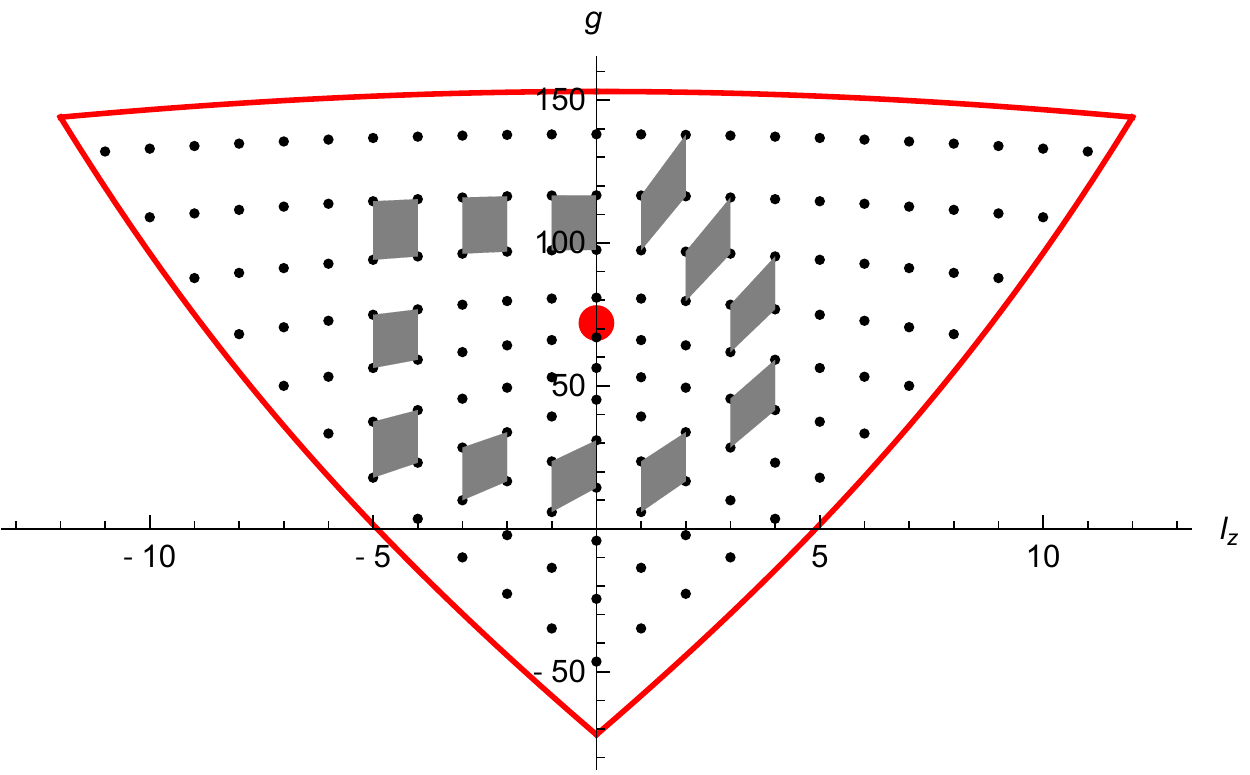}
\caption{Quantum lattice formed by  the joint spectrum (black dots) for the  commuting operators $(\hat L_z, \hat G)$ for
fixed principal quantum number $n=12$ and $a=144/5$, illustrating the monodromy 
of a fundamental cell transported around the isolated critical value of the classical energy momentum map (red dot). } \label{fig:Qmono}
\end{figure}

In this article we show that the spectrum of joint eigenvalues $(E,g,l_z)$ of $\mathcal{G}$ has quantum 
monodromy in the limit of sufficiently large principal quantum number $n$.
The joint spectrum of $\mathcal{G}$  for a large but fixed principal quantum number $n$ is shown in Fig.~\ref{fig:Qmono}. 
Locally the spectrum has a lattice structure. 
Globally however it has a defect as can be seen from moving a fundamental lattice cell about the isolated singular value 
of the energy momentum map marked by the red dot in Fig.~\ref{fig:Qmono}.  
As a consequence globally defined quantum numbers cannot exist.
Even though separability of the hydrogen atom in prolate spheroidal coordinates has been known for a long time,
this is the first time that quantum monodromy in the Hydrogen atom is described. 

{\em Super-integrable Systems.} 
Before we discuss the details of the Hydrogen atom we want to elucidate the general 
structure underlying our analysis.
Given a Hamiltonian operator $\hat H$ it is exceptional to be able to find a QIS
$\mathcal{H}$ that contains $\hat H$ -- most Hamiltonians are non-integrable.
It is even more exceptional to be able to find non-trivially distinct QIS 
$\mathcal{ H}$ and $\mathcal{G}$ that both contain the same $\hat H$.
This implies, but is not equivalent to, that the system is superintegrable, 
see, e.g., the review  \cite{Evans90}.
Important examples  are systems that are separable in different coordinate systems.
Schwarzschild \cite{Schwarzschild1916} %
was the first to point out that if the Hamilton-Jacobi equation of $H$ can be separated in more 
than one coordinate system, the quantum energy eigenvalues of $\hat H$ are degenerate.
Such a Hamiltonian operator $\hat H$ is called multiseparable, 
and is hence included in non-equivalent QIS's $\mathcal H$ and $\mathcal G$. 
The simplest multiseparable systems are the free particle, the harmonic oscillator, and the Kepler problem. 
A multiseparable system is superintegrable, because if both 
$\mathcal{H}$ and $\mathcal{G}$ contain $\hat H$, then we have found 
more than $f-1$ operators that commute with $\hat H$.
An  important  class of 3-dimensional superintegrable and multiseparable systems
is classified in \cite{Kress06}.

The classical geometry of superintegrable systems is well understood.  Fixing the integrals defines tori of lower dimension than in the Liouville-Arnold theorem and Nekhoroshev showed that one can construct lower dimensional action angle coordinates in a kind of generalization of the Liouville-Arnold theorem \cite{Nekhoroshev1972}. 
More global aspects have been studied in \cite{FomenkoMishchenko78,Delzant87}.
From the classical geometric point of view considering tori with half the dimension of phase space in 
a super-integrable system appears somewhat arbitrary. However, from the quantum point of view it 
is prudent to study all possible sets of commuting observables, because these tell us what can 
be measured simultaneously as the uncertainty principle is trivial in this case.
Thus we are going to study a particular set of collections of Kepler ellipses that form 3-tori in phase space, 
and we will show that the joint quantum spectrum associated to these tori has quantum monodromy.

{\em The Kepler problem and the hydrogen atom.}
To fix our notation let $\mathbf{r} = (x,y,z)^t$ be the position of the electron in $\mathbb{R}^3$
and $\mathbf{p} = (p_x, p_y, p_z)$ its momentum. 
The nucleus is at the origin. The Hamiltonian is 
$H = \tfrac12 |\mathbf{p}|^2 - 1/r$ where $r = |\mathbf{r}|$, the angular momentum is
${\mathbf L} = \mathbf{r} \times \mathbf{p}$,
and the Laplace-Runge-Lenz  vector is 
${\mathbf e} = {\mathbf p} \times \mathbf{L} - \mathbf{r}/r$.
The components of $\mathbf{L}$ and $\mathbf{e}$ all have vanishing Poisson brackets with $H$,
but are not all independent because of the relations 
$\mathbf{L} \cdot \mathbf{e} = 0$ and $|\mathbf{e}|^2 = 1 + 2 H |\mathbf{L}|^2$. 
Hence there are five independent  
integrals and the system is maximally superintegrable,  i.e.{}  fixing the five integrals on the six-dimensional phase space of the Kepler problem 
 for negative energies defines one-dimensional tori. These one-dimensional tori simply are the periodic orbits given by the Kepler ellipses. 
Introducing $\mathbf{K} = n \mathbf{e}$ where $H = - 1/(2 n^2)$ (assuming $H < 0$) the components
of $\mathbf{L}$ and $\mathbf{K}$ satisfy the commutator relations of the algebra  $so(4)$ 
with Casimirs $\mathbf{L} \cdot \mathbf{K} = 0$ and $\mathbf{L}^2 + \mathbf{K}^2 = n^2$.
In the quantum version similarly operators $\hat H$, $\hat{\mathbf{L}}$ and 
$\hat{\mathbf{K}}$ are defined and satisfy the same commutation relations 
of the $so(4)$ algebra. We note that the $so(4)$ symmetry was already used by Pauli in his 1926 paper \cite{Pauli1926} to determine the spectrum of hydrogen.


{\em Separation in Prolate Spheroidal Coordinates and Monodromy.}
Prolate spheroidal coordinates $(\xi,\eta,\phi)$ with focus points at $\pm \mathbf{a} = (0,0,\pm a)$ on the $z$-axis 
are defined by
$\xi = (r_1+r_2)/(2a)$, $\eta=(r_1-r_2)/(2a)$ where 
$r_1=\vert \mathbf{r} -   \mathbf{a}  \vert$ and $r_2=\vert \mathbf{r} +   \mathbf{a}  \vert$
and $\phi$ is the angle about the $z$-axis. 
The surfaces of constant $\xi$ and  $\eta$ are prolate ellipsoids and two-sheeted hyperboloids, respectively, with focus points $\pm \mathbf{a}$.
For $a\to 0$, spherical coordinates are recovered, 
while for $a\to\infty$, parabolic coordinates are found. 
Assuming the hydrogen nucleus to be located at the focus point $\mathbf{a}$,
the separation of the Schr\"odinger equation 
$(-\frac12\nabla^2-1/r_1)\psi = E\psi$ with the  ansatz
$\psi(\mathbf{r} )=\psi_\xi(\xi) \psi_\eta(\eta) \psi_\phi (\phi)$ leads to three separated equations.
The equation for $\phi$ gives the angular momentum eigenvalues $l_z=m=0,\pm1, \pm 2,\ldots$, 
and the equations for $\xi$ and $\eta$ give the same equation
\begin{equation}\label{eq:sepSchr}
-\frac{\ud }{\ud s} \left(s^2-1\right) \frac{\ud }{\ud s} \psi_s(s) = \frac{P(s)}{s^2-1} \psi_s(s),
\end{equation}
where $P(s)$ is the polynomial 
\begin{equation}\label{eq:defPs}
P(s)  =  \big( 2 a^2 E\, (s^2 -1) + 2 a s - g\big) (s^2-1)  - l_z^2. 
\end{equation}
For $s=\xi $, Eq.~\eqref{eq:sepSchr} is considered on the interval $[1,\infty)$ and for $s=\eta$,   Eq.~\eqref{eq:sepSchr}  is considered on  $[-1,1]$.
The separation constant $g$ is the eigenvalue of the operator
\begin{equation} \label{eq:defhatG}
\hat{G} = \hat{\mathbf{L}}^2 + 2 a\, \hat{e}_z,
\end{equation}
where 
$
\hat{e}_z  
$
is the $z$-component of the quantum Laplace-Runge-Lenz vector which in position representation reads 
\begin{equation}\label{eq:defhatez}
\hat{e} = \frac12 \big( \hat{\mathbf{p}}\times \hat{\mathbf{L}} - \hat{\mathbf{L}}\times \hat{\mathbf{p}}   \big) - \frac{1}{r_1} (\mathbf{r} - \mathbf{a} ).
\end{equation}
In \eqref{eq:defhatG} and \eqref{eq:defhatez}  $\hat{\mathbf{p}}=  -i \nabla$ and $ \hat{\mathbf{L}} = - i  (\mathbf{r} -\mathbf{a}) \times \nabla$ are the standard momentum and angular momentum operators defined relative to the point  $\mathbf{a}$. 
The operators $\hat{H}$, $ \hat{L}_z$ and $\hat{G}$  mutually commute, i.e.{}  ${\mathcal G} = (\hat{H}, \hat{G}, \hat{L}_z)$ defines a QIS. 



\begin{figure}
 \includegraphics[width=4cm]{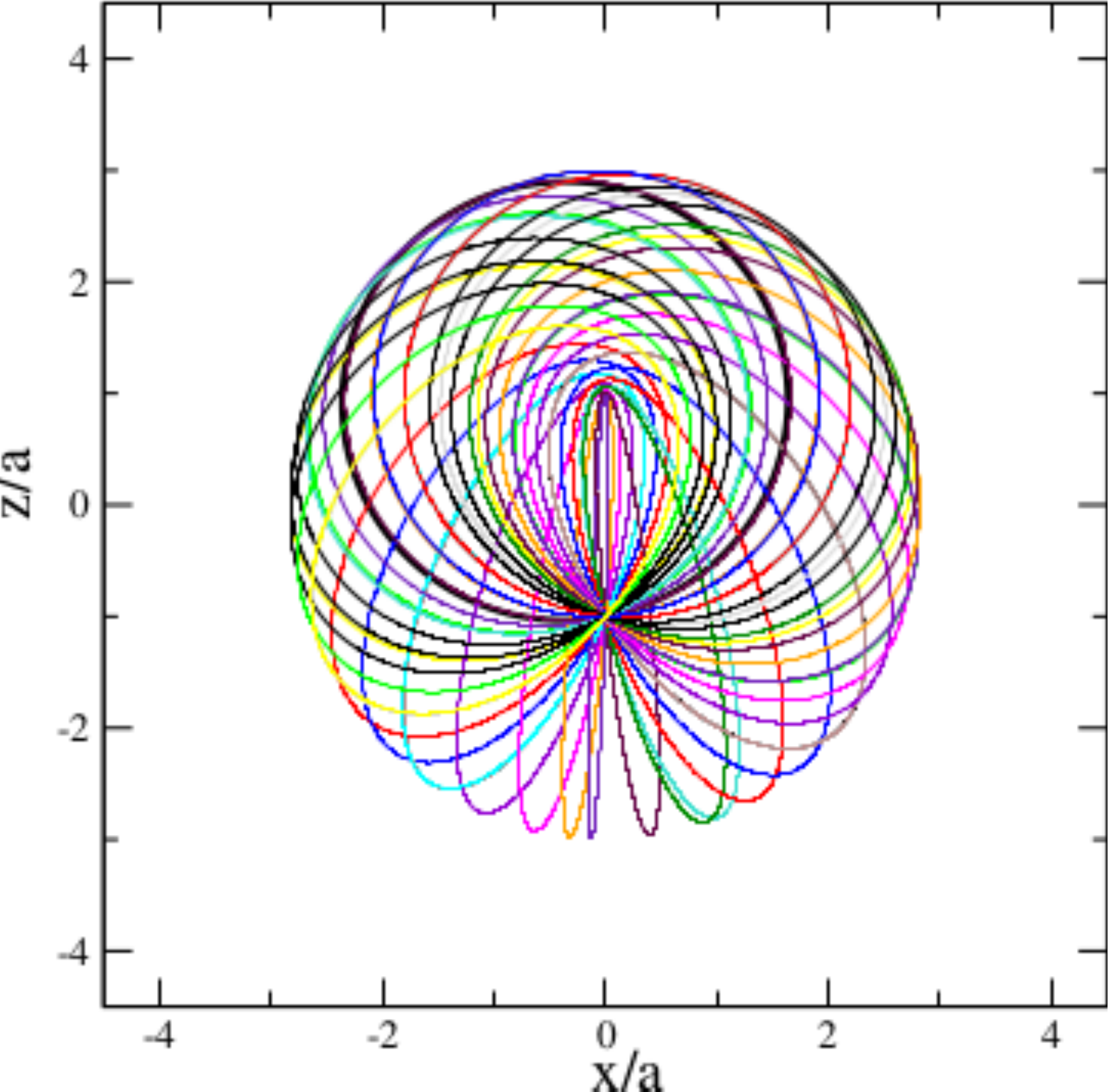}
 \includegraphics[width=4cm]{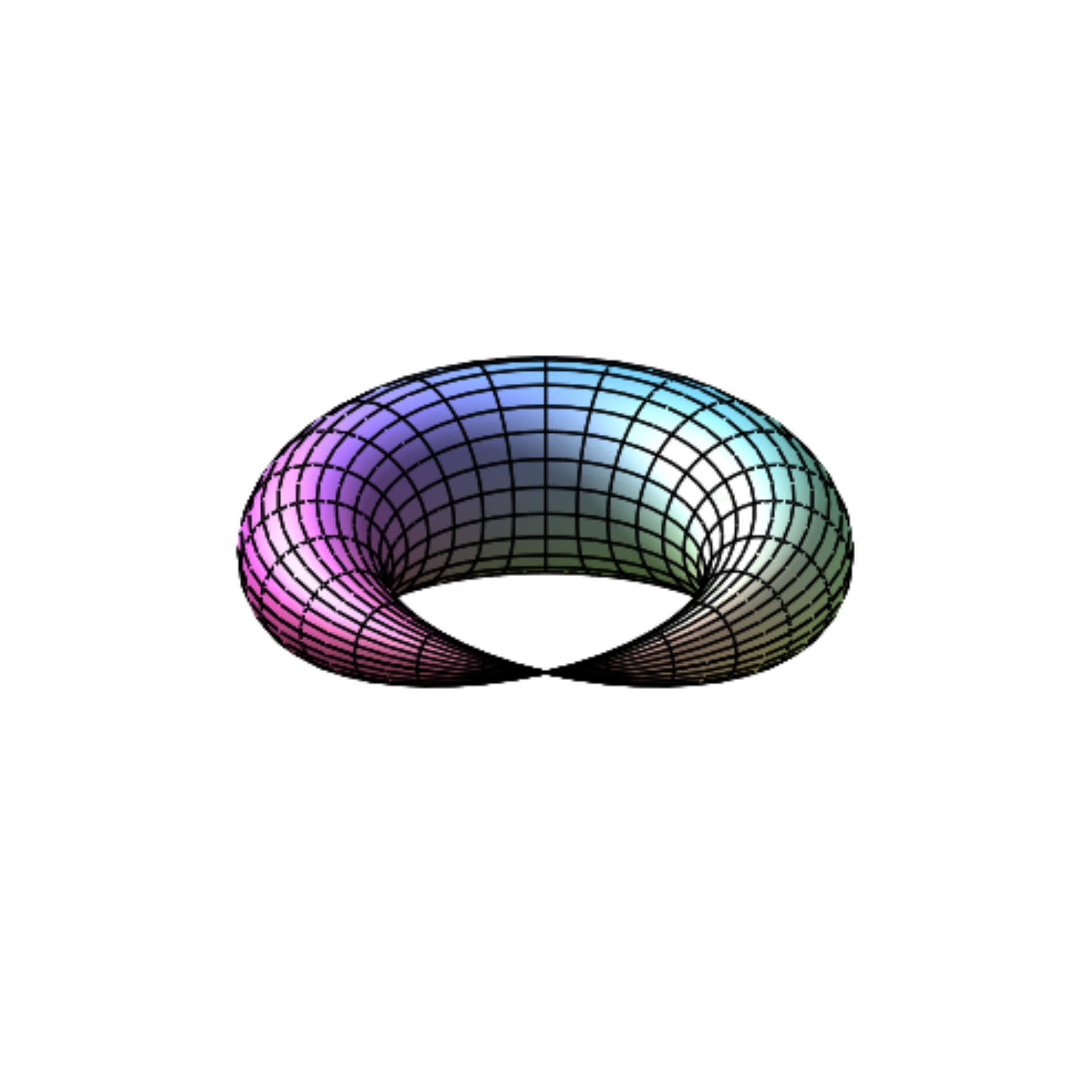}
\caption{Kepler ellipses corresponding to the isolated singular value of the energy momentum map (left) and pinched 2-torus (right).}
\label{fig:pinched}
\end{figure}

The joint spectrum of $ (\hat{H}, \hat{G}, \hat{L}_z)$ can be computed numerically using a shooting method for the system of Fuchsian equations formed with $s=\eta$ and $s=\xi$ in  \eqref{eq:sepSchr}.
We follow however a more efficient algebraic approach that  is briefly described in the methods section.   A standard WKB ansatz shows that the joint spectrum can be computed semi-classically from a Bohr-Sommerfeld quantization of the actions according to
$I_\phi =  \frac{1}{2\pi}  \oint p_\phi \,\ud \phi = l_z = m $,  
$I_\eta = \frac{1}{2\pi}  \oint p_\eta \,\ud \eta  = n_\eta+\frac12 $ and $I_\xi =  \frac{1}{2\pi}   \oint p_\xi \,\ud \xi = n_\xi + \frac12 $ with 
$m\in \Z$
 and non-negative quantum numbers $n_\eta$ and $n_\xi$. 
 Here the momenta $p_\eta$ and $p_\xi$ are given by
$ p_s^2 = P(s)/(s^2-1)$ where $P(s)$ is again the polynomial in Eq.~\eqref{eq:defPs} which implies that the actions $I_\eta$ and $I_\xi$ are given by elliptic integrals.
It turns out that $I_\eta$ and $I_\xi$ are not smooth  functions of the eigenvalues $(E,g,l_z)$ -- 
 from a study of the elliptic integrals in the complex plane
 it  can be shown that they have a discontinuous derivative  at $l_z=0$ (this can be seen as a special case of the computation in \cite{WaalkensDullinRichter04}).  This is an indication that the lattice of eigenvalues formed by the joint spectrum might have a defect. 
Using the calculus of residues one finds that the 
sum  of the actions $I_\eta + I_\xi + \vert  I_\phi \vert $ is equal to $1/\sqrt{-2E}$ and hence is in particular smooth for energies $E<0$. In fact the  quantization of the sum can be identified with the principal quantum number $n$ from which we then obtain $E=-1/(2 n^2)$. The magnetic quantum number $m$ and the principal quantum number $n$ are the only good quantum numbers -- no third quantum number can be globally defined for the QIS $\mathcal{G}$.
This is what we see in Fig.~\ref{fig:Qmono} which shows a layer of the lattice of the joint spectrum of constant principal quantum number $n$.  
The defect is caused by a singular value of co-dimension two of the classical energy momentum map
which is located on the $g$ axis at  $g= 2a$. 

\begin{figure}
\includegraphics[width=\columnwidth]{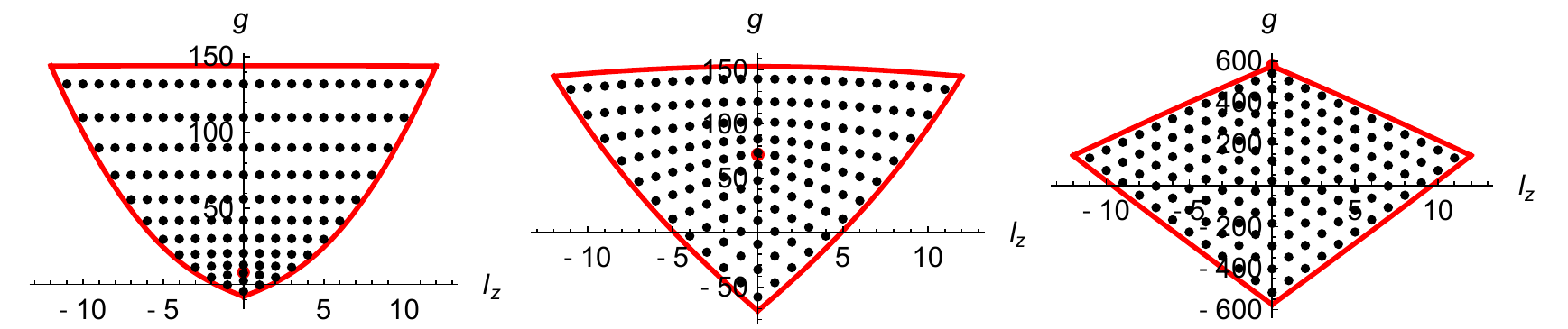}
\caption{Classical critical values (red) in  $(l_z, g)$, 
for $n=12$, $a = 4, 36, 288$ and corresponding joint spectrum (black dots). 
} \label{fig:MoMap}
\end{figure}

For $(E,g,l_z)$ at this singularity, the polynomial $P(s)$ in Eq.~\eqref{eq:defPs} has a double root at  $1$. Because of this, the separated classical motions in  the $\eta$ and $\xi$ degrees of freedom both have a turning point at $1$. The corresponding orbits in configuration space are given by a two-parameter family of Kepler ellipses 
which have a common mutual intersection point at the focus point $-\mathbf{a}$.  In the left panel of Fig.~\ref{fig:pinched} we show the one-parameter subfamily of these ellipses contained in the $(x,z)$ plane. The full two-parameter family is obtained from rotation about the $z$-axis. 
In phase space the two-parameter family topologically forms a  pinched 2-torus times a circle.  
In fact in order to prove monodromy  it is sufficient to show that the 
corresponding classical system has a pinched torus \cite{VuNgoc99} and this is what we do in the methods section.
The two-parameter family of these Kepler ellipses contains the degenerate ellipse consisting of the line segment given by the interval $[a+1/E,a]$ on the $z$-axis.  This is a  periodic collision orbit  that runs along the symmetry axis and bounces back and forth between the nucleus at $\mathbf{a}$ and the turning point $a+1/E$ on the $z$-axis. For $E$ greater than $-1/(2a)$ which is the value of the potential energy at the focus point $-\mathbf{a}$,  the turning point of this periodic orbit on the $z$-axis is below the focus point $-\mathbf{a}$. This is the condition for the isolated value for the energy momentum map to come into existence as we will show in the methods section. 
This means that the layers of the joint spectrum for constant  principal quantum number $n$ have a defect for $n > \sqrt{a}$ and no defect for $n<\sqrt{a}$, which is illustrated in Fig.~\ref{fig:MoMap}.

%

\vspace{1ex}
\centerline{\bf Conclusions}

We have shown that the joint spectrum associated with the separation of the hydrogen atom 
in prolate spheroidal coordinates has a lattice defect due to quantum monodromy.  
The quantum integrable system $\cal G$ obtained from this separation has two 
global quantum numbers $n$ and $m$, but a third global quantum number does 
not exist due to the lattice defect.
This raises the fascinating question whether an experiment can be designed which 
measures simultaneously the values of the the three observables of $\cal G$.

Spherical and parabolic coordinates can be considered to be limiting case of prolate ellipsoidal coordinates for $a\to 0$ and $a\to \infty$, respectively. In Fig.~\ref{fig:MoMap} the left and right images are close to these limiting situations.
For $a \to \infty$, the condition $n > \sqrt{a}$ cannot be satisfied, so there is no monodromy.
For $a \to 0$, the focus point collides with the boundary, and after extracting the square root of 
the diagram we recover the standard spectrum of $\cal H$ again without monodromy.
This leaves the sphero-conical coordinate system. Preliminary computations show that
this QIS does not have monodromy. However, the spectrum does have an unexpected 
structure in that there is an additional separatrix.

A similar type of analysis can be done in other multiseparable systems. 
In particular we have already obtained preliminary results for the isotropic harmonic
oscillator which again show that there is a QIS that contains this Hamiltonian for which 
the joint spectrum has monodromy. This will be presented in a forthcoming paper.

\vspace{1ex}
\centerline{\bf Methods}

\begin{figure}
\includegraphics[width=\columnwidth]{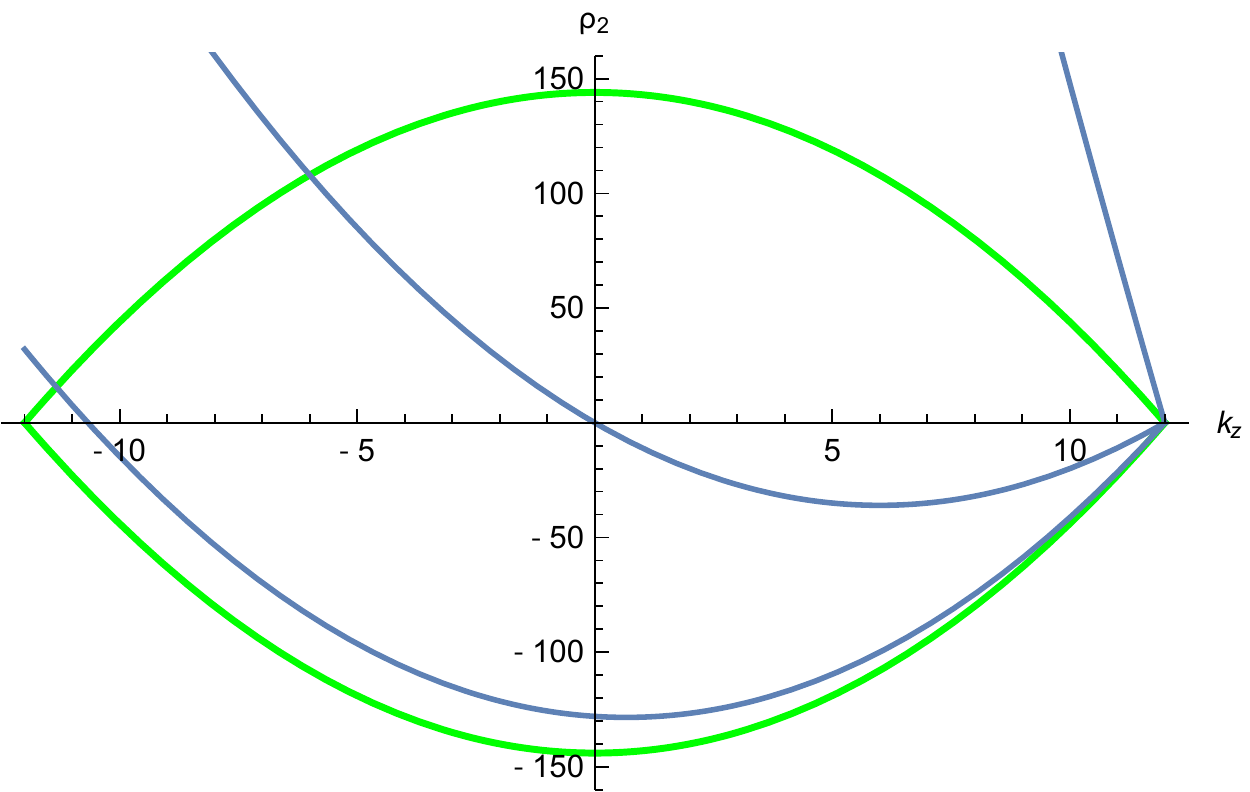}
\caption{Slice $\rho_3 = 0$ through the singular reduced phase space of the Kepler problem defined by the Casimir $C=0$
for $l_z  = 0$, $n=12$ and the lines $g = 2a$ for $a = 4, 36, 288$. The two smaller values of $a$ correspond to 
a pinched torus, the larger value of $a$ corresponds 
to an elliptic equilibrium point.}
\label{fig:Lemon}
\end{figure}

{\em Proof of monodromy. }
As mentioned in the main text we can prove the presence of monodromy by studying the analytic properties of the elliptic integrals that give the actions $(I_\phi , I_\eta, I_\xi)$ associated with the separation in prolate spheroidal coordinates.  It turns out  that these integrals do not have a smooth dependence on the values of the constants of motion $(E,g,l_z)$ and no smooth single valued  continuation of the actions is possible.   
The computations are very similar to the ones for Euler's two-center problem in  \cite{WaalkensDullinRichter04}.

Instead we here prove the presence of monodromy by showing the existence of a pinched torus using  the  method of \emph{singular reduction} \cite{CushBates}. 
Following  \cite{VuNgoc99} the  existence of a pinched torus implies quantum monodromy. 
In the present case we can follow an approach similar to 
the reduction of Stark-Zeeman perturbations of the Kepler problem 
that have been analysed in detail in \cite{CushmanSadovskii99,EfstathiouSadovskiiZhilinskii2007}.
The first step is then symplectic reduction of the periodic Hamiltonian flow of the Kepler problem which 
leads to a four-dimensional phase space given by the product of two spheres, $S^2 \times S^2$,  where the two spheres  have constant energy
$E = -1/(2n^2)$ and are given by
$(\mathbf{K} + \mathbf{L})^2 = n^2$ 
and 
$(\mathbf{K} - \mathbf{L})^2 = n^2$.
On this reduced space we then have an integrable Hamiltonian system with integrals 
$G$ and $L_z$.  
The essential difference to  \cite{CushmanSadovskii99,EfstathiouSadovskiiZhilinskii2007}  is that our 
integral $G$ does not originate from a perturbation in their class.
Reduction of the axial rotational symmetry generated by $L_z$ leads to a 
singular one-degree-of-freedom system which can be described using a 
Poisson structure in $\mathbb{R}^3$ with variables 
$\rho_1 = K_z$, $\rho_2 = L_x^2 + L_y^2 - K_x^2 - K_y^2$ and $\rho_3 = K_x L_y - K_y L_x$. The variables have
Poisson brackets 
$\{ \rho_1, \rho_2 \} =  2 \rho_3 $, 
$\{ \rho_1, \rho_3 \} = -2 \rho_2 $, 
$\{ \rho_2, \rho_3 \} = 4 \rho_1 ( n^2 + m^2 - \rho_1^2)$
with the Casimir
$C = ( n^2 - (m + \rho_1)^2)(n^2 - ( m - \rho_1)^2) - \rho_2^2 - \rho_3^2 = 0$
where $m = l_z$ is a parameter  \cite{CushmanSadovskii99}.
Our reduced second integral is $G = \frac12( \rho_2 + (n^2 - \rho_1^2) + m^2 ) + 2 a \rho_1 / n$.
The two-dimensional phase space defined by $C=0$ and $|K_z| \le n$ is singular when $m = l_z = 0$
with singular points at $(\pm n, 0,0)$,  see Fig.~\ref{fig:Lemon}. 
For $|m| < n$ the reduced phase space is diffeomorphic to a sphere,
for $|m| = n$ it is a point at the origin and empty otherwise. 
When the surface $G = g$ intersects the surface $C=0$ in the singular point
the system has an equilibrium point. When the intersection is otherwise empty 
it is an elliptic equilibrium. When the intersection contains non-singular points 
then we have a pinched torus and hence monodromy.
To see when the latter case occurs we compare the slope of the Casimir surface 
at the singular point which is $n$ with the slope of $g=2a$ at that point, 
which is $2n - 4a/n$, so that a bifurcation occurs for $a = n^2$, see Fig.~\ref{fig:Lemon}.. 
For $a \ge n^2$, there is no monodromy, while for $0< a  < n^2$ there is. This is demonstrated in Fig.~\ref{fig:MoMap}.

\begin{figure}
\includegraphics[width=\columnwidth]{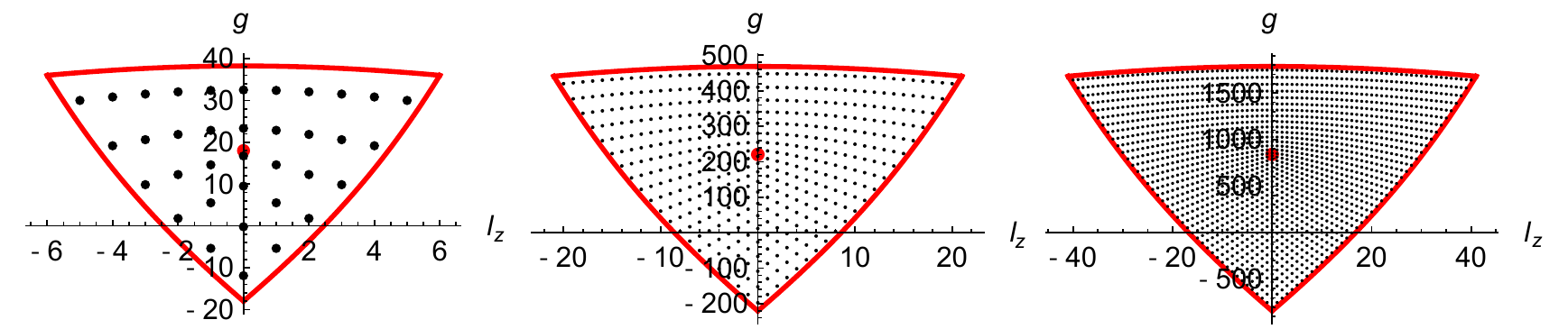}
\caption{Approaching the semiclassical limit $n = 6, 21, 41$, $a = n^2/4$.} \label{fig:MoMapSemi}
\end{figure}

The proof that the presence of a pinched torus implies quantum monodromy requires semiclassical techniques~\cite{VuNgoc99}. 
For illustration, we show how the semiclassical limit develops
by increasing the principal quantum number $n$  in Fig.~\ref{fig:MoMapSemi}.  
From the Bohr-Sommerfeld quantization rule it follows that 
the locally smooth lines 
resulting from the spectrum in the semiclassical limit  can be identified with the contours of the action functions. From Fig.~\ref{fig:MoMapSemi}
it can be inferred that smooth actions are necessarily multi-valued.

{\em Computation of the spectrum}.
The computation of the joint spectrum of the QIS  $\mathcal{G} = (\hat H, \hat G, \hat L_z)$ 
obtained from separating the Hamiltonian $\hat{H}$
of the  Hydrogen atom in prolate spheroidal coordinates is most 
efficiently done using interbasis expansion.
This has been discussed by Coulson \& Robinson \cite{CoulsonRobinson1958}.
Using an adaption of the separation in spherical coordinates they show that
 the spectrum can be computed  exactly from eigenvalues of 
matrices $A$ of size $(n - |m|) \times (n-|m|)$. 
The matrix $A$ is tri-diagonal and symmetric, 
the entries are
$A_{l,l} = l(l+1)$ and 
$A_{l+1,l} =  \frac{a}{n}  \sqrt{    (n^2 - l^2) ( l^2 - m^2)   / (  l^2 - \tfrac14) }$
where $l = \vert m \vert, \vert m \vert +1,\ldots, n-1 $.
%
For fixed $n$ and $m$, this produces one vertical line of the joint spectrum shown 
in the figures.
As discussed in \cite{CoulsonRobinson1958}, it is very special to have spheroidal wave functions
that can be written as a finite combination of spherical harmonics. 
For the related Euler's two-centre problem,  no such representation is possible \cite{Coulson1967}.


%

\end{document}